\documentclass[letterpaper, 10 pt, conference]{ieeeconf}  

\IEEEoverridecommandlockouts                              

\usepackage[utf8]{inputenc}
\usepackage[style=ieee]{biblatex}

\usepackage{siunitx}
\usepackage[super]{nth}
\usepackage{amsmath,amssymb,amsfonts, mathtools}
\usepackage{hyperref} 
\usepackage{physics}
\usepackage{algorithmic}
\usepackage{graphicx}
\usepackage{stfloats}
\graphicspath{Figures/}
\usepackage{textcomp}
\usepackage{xcolor}
\usepackage{soul}
\let\labelindent\relax
\usepackage{enumitem}
\usepackage[ruled]{algorithm2e}
\def\BibTeX{{\rm B\kern-.05em{\sc i\kern-.025em b}\kern-.08em
    T\kern-.1667em\lower.7ex\hbox{E}\kern-.125emX}}
\DeclareMathOperator*{\argmax}{arg\,max} 
\def\thesubsubsectiondis{\unskip\arabic{subsubsection})}

\usepackage{tikz}

\newcommand\copyrighttext{%
  \footnotesize \textcopyright 2023 IEEE. Personal use of this material is permitted. Permission from IEEE must be obtained for all other uses, including reprinting/republishing this material for advertising or promotional purposes, collecting new collected works for resale or redistribution to servers or lists, or reuse of any copyrighted component of this work in other works.}

\newcommand\copyrightnotice{%
\begin{tikzpicture}[remember picture,overlay]
\node[anchor=south,yshift=10pt] at (current page.south) {\fbox{\parbox{\dimexpr0.75\textwidth-\fboxsep-\fboxrule\relax}{\copyrighttext}}};
\end{tikzpicture}%
}

\renewcommand\fbox{\fcolorbox{red}{white}}
\begin{document}

\title{Decision-change Informed Rejection Improves Robustness in Pattern Recognition-based Myoelectric Control}

\author{Shriram Tallam Puranam Raghu$^{1}$, Dawn MacIsaac$^{1}$, and Erik Scheme$^{1}$,\IEEEmembership{~Senior~Member,~IEEE}
\thanks{*This work was supported in part by NSERC Grant 2020-04776, Canada}
\thanks{Shriram Tallam Puranam Raghu, Dawn MacIsaac, and Erik Scheme are with the Department of Electrical and Computer Engineering and the Institute of Biomedical Engineering, University of New Brunswick, Fredericton, NB, E3B 5A3, Canada. 
        {Email: \tt\footnotesize stallam@unb.ca, dmac@unb.ca, escheme@unb.ca}}%
}

\maketitle

\copyrightnotice

\begin{abstract}

Post-processing techniques have been shown to improve the quality of the decision stream generated by classifiers used in pattern-recognition-based myoelectric control. However, these techniques have largely been tested individually and on well-behaved, stationary data, failing to fully evaluate their trade-offs between smoothing and latency during dynamic use. Correspondingly, in this work, we survey and compare 8 different post-processing and decision stream improvement schemes in the context of continuous and dynamic class transitions: majority vote, Bayesian fusion, onset locking, outlier detection, confidence-based rejection, confidence scaling, prior adjustment, and adaptive windowing. We then propose two new temporally aware post-processing schemes that use changes in the decision and confidence streams to better reject uncertain decisions. Our decision-change informed rejection (DCIR) approach outperforms existing schemes during both steady-state and transitions based on error rates and decision stream volatility whether using conventional or deep classifiers. These results suggest that added robustness can be gained by appropriately leveraging temporal context in myoelectric control.

\end{abstract}
Keywords - Surface Electromyography, sEMG, Pattern Recognition, myoelectric control, steady-state, transitions, temporal, post-processing, majority vote, rejection, Bayesian fusion, outlier detection, DCIR, VoCIR.

\pagestyle{plain} 

\section{Introduction}

Surface Electromyography (sEMG) signals contain rich information about user intent, and as such, have been studied extensively in fields such as prosthesis control \cite{asghari_oskoei_myoelectric_2007}, gesture recognition \cite{chen_pattern_2013} and rehabilitation \cite{ho_emg-driven_2011}. Several new techniques and improvements have been proposed for the sEMG Pattern Recognition (PR) pipeline including novel features \cite{al-timemy_improving_2016}, deep Learning (DL) classifier models \cite{zhai_self-recalibrating_2017}, and pre-processing techniques \cite{geethanjali_comparative_2015}. However, factors such as variation in contraction intensity and limb position still degrade the performance of PR systems and manifest as errors in the resulting classification decision stream \cite{campbell_current_2020}. This has motivated researchers to propose a class of algorithms that use additional context (e.g. confidence stream) to increase the accuracy and stability of the decision stream. These algorithms can broadly be grouped together as Decision Stream Quality Improvement (DSQI) algorithms. DSQI techniques include, for example, post-processing algorithms \cite{khushaba_toward_2012}, adaptive classifiers \cite{campbell_linear_2019}, and outlier rejection \cite{ding_adaptive_2019}. 

Although such techniques have been shown to improve accuracy, it has also been reported that improvements in accuracy do not necessarily translate to improvements in online usability \cite{scheme_electromyogram_2011}. As such, a usability test such as a Fitt's law test is often encouraged to assess the impact of algorithm changes on user-in-the-loop performance \cite{wurth_real-time_2014}. Nevertheless, offline evaluation remains widely in use because it is more convenient and reproducible, and enables the direct comparison of approaches on the same dataset.

In a previous study \cite{tallam_puranam_raghu_analyzing_2022}, we highlighted that the lack of correlation between offline accuracy and usability may be due to the controlled conditions for offline tests, which do not represent the real-world device usage \cite{chang_wearable_2020}. Most offline studies collect a set of separate motions (steady-state or ramp contractions) that are used to train and then test the sEMG-PR algorithms. Such constrained protocols simplify the labeling process, but do not reflect the feed-forward use of sEMG-PR devices, which necessarily involves dynamic transitions between various contractions. Thus, the accuracies reported by offline studies provide an incomplete picture of the overall performance of the sEMG-PR system. In \cite{tallam_puranam_raghu_analyzing_2022} we proposed a novel framework consisting of several metrics that assess classifier performance during transitions in addition to steady-state to bridge this gap in offline evaluation. Using this framework, we showed that the behaviour of classifiers may differ substantially during transitions even if their steady-state performance is comparable.

In this work, we build on our previous work by using the proposed framework to study the impact of various DSQI algorithms on offline performance and use dynamic data containing transitions. Several DSQI algorithms have been proposed in the literature, but none have been fully evaluated against each other, nor have they been evaluated in the context of data containing transitions. %
Consequently, this work makes the following contributions:
\begin{enumerate}   
    \item A review of established DSQI algorithms with a head-to-head comparison.
    \item Two novel techniques that \emph{adapt and respond} to transitionary behaviours.
    \item An evaluation of these algorithms on a new large dynamic dataset using both steady-state and transition metrics.
\end{enumerate}
\hfill \hfill

\section{Review of Established DSQI Algorithms}

\begin{figure*}[htbp]
    \centering
    \includegraphics[width=0.9\textwidth]{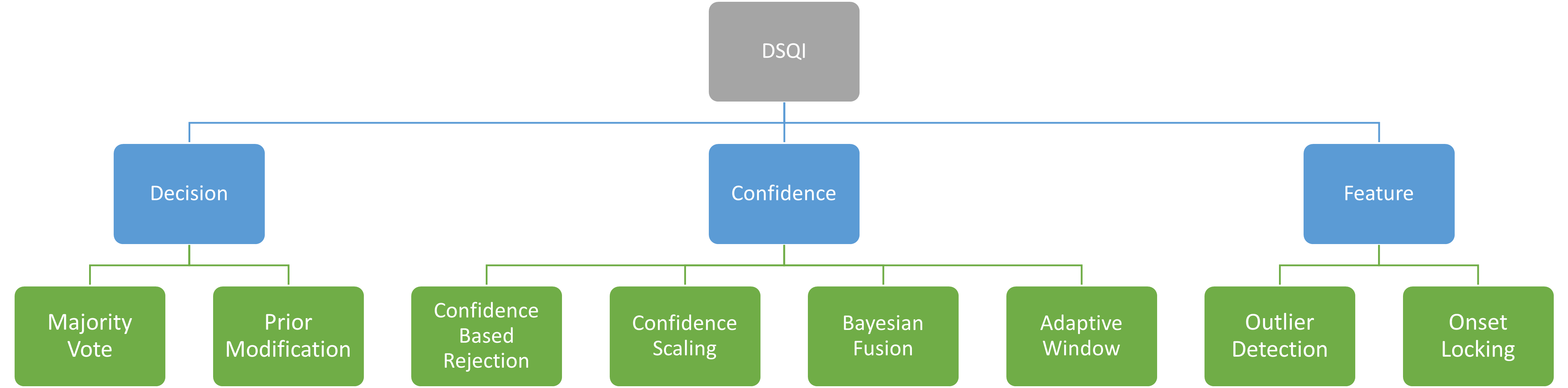}
    \caption{Decision Stream Quality Improvement (DSQI) schemes grouped by the main source of information used.}
    \label{fig:DSQI_Chart}
\end{figure*}

Various DSQI techniques have been reported in the literature with some success. For instance, post-processing rejection is well-established in the context of dynamic movements \cite{scheme_confidence-based_2013, robertson_effects_2019}, and is widely used \cite{Bao_Zaidi_Xie_Yang_Zhang_2022, Krasoulis_Vijayakumar_Nazarpour_2020}. Established DSQI schemes can be grouped into three categories based on the primary source of information being used: Feature-Based Schemes, Confidence-Based Schemes, and Decision-Based Schemes. An overview of this taxonomy is shown in Figure \ref{fig:DSQI_Chart}. Note that most rejection schemes are confidence-based because they rely on the confidence output from the classifier. In order to delineate the schemes, we first define some common terminology. Let $\mathrm{EMG} \in \mathbb{R}^{N_S \times N_{\mathrm{CH}}}$ be the EMG dataset containing $N_S$ samples from $N_{\mathrm{CH}}$ channels. Let $\mathrm{EMG}_{i} \in \mathbb{R}^{M \times N_{\mathrm{CH}}}$ be the $i\textsuperscript{th}$ EMG frame which is a matrix of size $M \times N_{\mathrm{CH}}$ with $M = \mathrm{Frame\ Length} \times \mathrm{Sample\ Rate}$ denoting the number of samples in a frame. Then we can consider $\mathrm{EMG}_{i}$ to be a collection of signals across channels, i.e. $\mathrm{EMG}_{i} = [S_{i1}, S_{i2}, \ldots, S_{iN_{\mathrm{CH}}}]$, $S_{ij} \in \mathbb{R}^{M}$. Let $h()$ denote the feature extractor function that takes in $\mathrm{EMG}_{i}$ and returns the corresponding feature vector $X_i \in \mathbb{R}^d$. Assuming that there are $N$ frames and $K$ classes, let the corresponding supervised labels be denoted by $Y = [y_1, y_2, \ldots, y_N]$ where $y_i \in \{1, \ldots, K\}, i = 1, \ldots, N$. $X_i$ is the input to an arbitrary classifier $f()$ which outputs a vector of confidences $C_{i} = [c_{i1}, c_{i2}, \ldots, c_{iK}]$ for frame $i$ which corresponds to confidence values of the $k = 1, \ldots, K$ classes. Then by definition:
\begin{gather}    
    X_i = h(\mathrm{EMG}_{i}) \label{eqn:feat_extract}\\
    C_i = f(X_i) \label{eqn:class_func}\\
    0 \leq c_{ik} \leq 1 \label{eqn:conf_range} \\
    c^{\mathrm{sum}}_{i} = \sum_{k = 1}^{K} c_{ik} = 1 \label{eqn:conf_sum}\\
    \hat{y}_i = \argmax_k (c_{ik}) \label{eqn:class_dec}\\
    \check{c}_i = \max_k (c_{ik}) \label{eqn:class_max_conf}
\end{gather}
\noindent where $\check{c}_i$ is the highest confidence for frame $i$ across all $K$ classes, and $\hat{y}_i \in \{1, \ldots, K\}$ is the class decision corresponding to the maximum confidence value. 

Having established this terminology, the various DSQI algorithms are described below, and summarized in Table \ref{tbl:summary_dsqi_schemes}, categorized by decision-based, confidence-based, and feature-based schemes.

\hfill \break
\noindent \emph{Decision-based Schemes}: Decision-based schemes use the decision outputs from a classifier to smooth out the decision stream. Though classifiers in PR-based myoelectric control traditionally treat each frame as being independent from each other, there is an inherent temporal ordering in the frames. The temporal evolution of the frames belonging to a given contraction is governed by the dynamics of the movement, which provides context that can be used to further improve the quality of the decision stream. Two common schemes in this group are:\\

    \begin{itemize}[leftmargin=*]
        \item \emph{Majority Vote}: Majority Vote (MV) \cite{englehart_robust_2003} is a simple but popular scheme that combines the current decision $\hat{y}_i$ and past $m$ decisions ($\hat{y}_{i-1}, \hat{y}_{i-2}, \ldots, \hat{y}_{i-m}$) [for a total window length of $m+1$] to re-estimate the new decision $\tilde{y}_i$ as given by:
        \begin{equation}
            \tilde{y}_i = \mathrm{mode}(\hat{y}_i, \hat{y}_{i-1}, \hat{y}_{i-2}, \ldots, \hat{y}_{i-m})
        \end{equation}
        
        where $\mathrm{mode}(.)$ returns the most frequent value. MV is simple to implement and effectively removes small blips in the decision stream at the cost of an introduced lag that is proportional to the MV length, $m$. \\
        
        \item \emph{Prior-Adjustment LDA}: In prior-Adjustment Linear Discriminant Analysis (pLDA) \cite{campbell_linear_2019}, instead of assuming equal prior probabilities across classes (as is typically the case) the class priors are dynamically varied by a supervisory system after each decision, favouring recently observed decisions. The pLDA update rule is as follows:    
        \begin{equation}
        \begin{gathered}
            P(k|\hat{y}_{i + 1}) = 
            \begin{cases*}
                P^+(k|\hat{y}_{i}), & $P^+(k|\hat{y}_{i}) < P_{\max}$\\
                P(k|\hat{y}_{i}) & o.w.
            \end{cases*}\\
            P^+(k|\hat{y}_{i}) = P(k|\hat{y}_{i}) + \Delta_n\\
            \Delta_n = b^s
        \end{gathered}
        \end{equation}
        where $P(k|\hat{y}_{i})$ represents the prior probability of observing class $k$ given decision $\hat{y}_i$, $b$ represents the growth rate, $s$ represents the number of consecutive decisions of the same class, and $P_{\max}$ represents the maximum permitted prior ($0 < P_{\max} < 1$). pLDA is a simple tweak to the Linear Discriminant Analysis (LDA) classifier and as such, it is easy to implement with Bayesian-type classifiers. This technique, however, is difficult to use with models like SVM that do not capture the underlying distribution of the classes, which is required for Bayesian inference.
        
    \end{itemize}
    
\noindent \emph{Confidence-based Schemes}: Confidence-based schemes are designed to utilise the confidence stream that a classifier can output, in addition to the decisions. The confidence values can be used to accept a decision, or change it to smooth the decision stream. Four popular schemes in this group are:\\

\begin{itemize}[leftmargin=*]
    \item \emph{Confidence Based Rejection}: Confidence-based rejection (CBR) \cite{scheme_confidence-based_2013} involves overwriting class decisions with low confidences to a different class. This relies on the classifiers themselves being able to provide a measure of the confidence of the output class (e.g. via posterior probabilities) that can be used to gauge if the decision is to be accepted or rejected. CBR is given by:        
    \begin{equation}
            \tilde{y_i} = \begin{cases*}
                \hat{y}_i & if $\check{c}_i > \mathrm{Th_{Rej}}$,\\
                \bigotimes & o.w.
            \end{cases*}
    \end{equation}        
    where $\mathrm{Th_{Rej}}$ is called the `Rejection Threshold' and $\bigotimes$ is the class that the decision will be assigned to if the class confidence is below the rejection threshold. Rejection is useful to mitigate unintended low confidence movements that degrade usability. However, the performance of rejection depends on the confidence distribution of the classifiers \cite{scheme_comparison_2015} , and the selection of an appropriate threshold. In this work, we reject low confidence decisions to the No Motion (NM) class ($k_\mathrm{NM}$).\\ 
        
    \item \emph{Confidence Scaling}: Confidence Scaling (CS) does not assume uniform class risk and correspondingly scales the class confidences $c_{ik}$, enabling the emphasis of certain classes more than others \cite{campbell_linear_2019}. The original study used this to emphasize the NM class, which results in the scheme behaving similar to CBR since NM class is emphasized during periods of low confidences, potentially resulting in the output being overridden to the NM class. The scheme in the original work was labelled `Cost Modification' and was based on the LDA, but the scaling can be trivially extended to any classifier. The new confidences $\tilde{c_{ik}}$ are given by:
        \begin{equation}
            \tilde{c}_{ik} = \delta \times c_{ik} \times s_k
        \end{equation}
        
    where $\delta$ is a constant used to scale $\tilde{c}_{ik}$ so that Eq. \ref{eqn:conf_sum} holds true and $s_k$ is a scaling factor associated with class $k$. The corresponding class decision $\tilde{y}_i$ is determined as in Eq. \ref{eqn:class_dec}. Similar to CBR, this emphasizes the NM class, suppressing low confidence decisions, but relies on having a good confidence distribution. If a classifier is over-confident in its decision, then the scheme may have minimal impact. \\
        
    \item \emph{Bayesian Fusion}: Bayesian Fusion (BF) \cite{khushaba_toward_2012} is similar to MV, but operates on confidences rather than the decision itself. A flaw in MV is that decisions with low and high confidences are given equal weighting, which is undesirable as low confidence (and potentially incorrect decisions) significantly impact the output decisions. BF addresses this issue by assigning weights that are proportional to the confidence of a decision and how temporally separated a decision is from the current one. More specifically, it uses the current confidence values $c_{ik}$ and past $m$ class confidences ($c_{(i-1)k}, c_{(i-2)k}, \ldots, c_{(i-m)k}$) [for a total window length of $m+1$] to re-estimate the new class confidences $\tilde{c}_{ik}$ and is given by:
        \begin{equation}
            \tilde{c}_{ik} = \delta \prod_{n=0}^{m} (c_{(i-n)k} + a_n)
        \end{equation}
        
        where $\delta$ is a constant to scale $\tilde{c}_{ik}$ so that Eq. \ref{eqn:conf_sum} holds true, and $a_n$ is a constant given by:        
        \begin{equation}
            a_n = 10 \times \frac{\exp(-0.5 \times \frac{n+1}{m+1})}{\sum_{l=1}^{m+1} \exp(-0.5 \times \frac{l}{m+1})}
        \end{equation}
        
        The corresponding class decision $\tilde{y}_i$ is determined as in Eq. \ref{eqn:class_dec}. Similar to MV, BF trades-off lag for smoothness, however the lag may be less pronounced than MV due to BFs weighting process.\\ 

    \item \emph{Adaptive Windowing}: Adaptive Windowing (AW) is based on the fact that, while shorter frames are desirable to improve responsiveness, longer frames provide more stable estimates and thus higher accuracy \cite{al-timemy_adaptive_2018}. Instead of holding the frame length fixed, adaptive windowing allows the frame length to be varied dynamically based on the confidence. Initially, features are extracted for a small frame length. If the confidence corresponding to the class decision for that frame is less than a threshold ($\mathrm{Th_{AW}}$), the scheme simply increases the frame length and reattempts the classification. This is repeated until either the confidence is high enough, or a maximum frame length is reached.
    Every time the confidence is ascertained to be low, the PR system returns the NM class ($k_\mathrm{NM}$) and increases the frame length. This has the effect of appending new EMG samples to the existing EMG frame and reclassifying the new data. If the max frame length is reached but the confidence is still low, $k_\mathrm{NM}$ is returned until the target confidence is reached. The scheme therefore acts similar to CBR. The frame length is reset to the base value once the target confidence is reached. The scheme is delineated in Algorithm \ref{alg:aw_scheme}.
        
        \SetKwComment{Comment}{/* }{ */}
        
        \begin{algorithm}
        \small
        \caption{Adaptive Window Algorithm}\label{alg:aw_scheme}
        $\mathrm{FL} \gets \SI{160}{\milli\second}$ \Comment*[r]{Frame Length}
        $\mathrm{FI} \gets \SI{16}{\milli\second}$ \Comment*[r]{Frame Increment}
        $\mathrm{ML} \gets \SI{256}{\milli\second}$ \Comment*[r]{Max Frame Length}
        
        \tcp{While new samples are available}
        \While{$\mathrm{EMG}$}{ 
            \tcc{Generate a frame from the latest samples}
            $\mathrm{EMG}_{i} \gets w(\mathrm{EMG}, \mathrm{FL}, \mathrm{FI})$\; 
            $X_i \gets h(\mathrm{EMG}_{i})$\;
            $\check{c}_i \gets f(X_i)$\;
          \eIf{$\check{c}_i \geq \mathrm{Th_{AW}}$}{
            $\tilde{y_i} \gets \hat{y}_i$\;
            $\mathrm{FL} \gets \SI{160}{\milli\second}$\;
          }{\If{$\mathrm{FL} < \mathrm{ML}$}{
              $\mathrm{FL} \gets \mathrm{FL} + \mathrm{FI}$\;
            }
            $\tilde{y_i} \gets k_\mathrm{NM}$\;
          }
        }
        \end{algorithm}
        
    AW may help mitigate errors due to smaller window sizes, but introduces computational burden as features from multiple window lengths must be extracted for training classifiers. Additionally, AW was designed with conventional classifiers (such as the LDA) in mind, but interactions with temporal classifiers such as the Long Short-Term Memory (LSTM), which requires both current and past frame values, remain unclear.\\ 
    
   \end{itemize}
    
\noindent \emph{Feature-based Schemes}: Feature-based schemes are designed to utilize information directly from the stream of feature vectors $X_i$. Although the classifiers are served these same features, discrete categorization during classification may lead to missed context in the decision stream. This context can be re-introduced in post-processing.
    
    \begin{itemize}[leftmargin=*]
    
    \item \emph{Onset Locking}: Onset Locking (OL) uses amplitude information to detect the onset and offset of movements to `lock' the decision stream of the PR system \cite{zhang_improving_2017}. This relies on that assumption that low amplitude values indicate a lack of active contraction, and therefore all decisions during this period can be locked to NM class ($k_\mathrm{NM}$). If a period of low amplitude is followed by a period of high amplitudes, then an onset of movement is assumed. Frames with high amplitudes values are then locked to the MV of the initial few decisions made by the classifier. In this work, we used the mean of the Mean Absolute Values (MAV) as the measure of activity, as given by:
        
        \begin{equation}
            \tilde{y_i} = \begin{cases*}
                k_{NM} & if $\overline{\mathrm{MAV}}_i < \mathrm{Th_{MAV}}$,\\
                \mathrm{LD}_i & o.w.
            \end{cases*}
            \label{eqn:locking_rej}
        \end{equation}
        
        where $\overline{\mathrm{MAV}}_i \in \mathbb{R}$ is the proportional control signal and $\mathrm{Th_{MAV}}$ is the threshold used to detect onset of movement. Let $\mathrm{MAV}_{ij} \in \mathbb{R}$ represent the MAV of channel $j$ for frame $i$ and let $\mathrm{MAV}_{i} \in \mathbb{R}^{N_\mathrm{CH}}$ represent a vector of MAVs across all channels, then $\overline{\mathrm{MAV}}_i$ is determined as:
        
        \begin{gather*}
            \mathrm{MAV}_{ij} = \mu(\abs{\mathrm{S_{ij}}})\\
            \mathrm{MAV}_i = [\mathrm{MAV}_{i1}, \ldots, \mathrm{MAV}_{iJ}]\\
            \overline{\mathrm{MAV}}_i = \mu(\mathrm{MAV}_i)
        \end{gather*}
                
        where $\mu(.)$ returns the mean. $\mathrm{Th_{MAV}}$ is determined using the NM class proportional control values across all the frames in the training data:
        
        \begin{gather*}
            \overline{\mathrm{MAV}}_\mathrm{NM} = [\overline{\mathrm{MAV}}_{i} \mid k = k_{\mathrm{NM}}]\\
            \mathrm{Th_{MAV}} = \mu(\overline{\mathrm{MAV}}_\mathrm{NM}) + 3\times \sigma(\overline{\mathrm{MAV}}_\mathrm{NM})
        \end{gather*}
         where $\sigma(.)$ returns the standard deviation. $\mathrm{LD}_i$ in Eq. \ref{eqn:locking_rej} is the locked decision calculated from the moment of onset. If the onset of movement is detected at frame $l$, then $L_i$ is determined by:
        
        \begin{equation}
            \mathrm{LD}_i = \begin{cases*}
                \hat{y}_i & if $i-l < m$,\\
                \mathrm{MV}(\hat{y}_l, \ldots, \hat{y}_{l+m-1}) & o.w.
            \end{cases*}
        \end{equation}
        
        That is, $\mathrm{LD}_i$ is locked to the MV of the first $m$ decisions from the onset of movement at frame $l$. The decision is locked until an offset is detected, at which point the process is reset until the next onset of movement is detected. OL may suppress spurious changes in the decision stream due to the locking behaviour, but necessitates a transition to NM between subsequent classes.\\ 
        
        \item \emph{Outlier Detection}: Outlier Detection (OD) attempts to reject decisions based on the deviation of features in the test data from distributions observed during training. Since most classifiers in PR systems are trained as multi-class classifiers, they inherently assign one of the known class labels to all incoming frames. This can lead the classifiers to incorrectly assign labels to outlier data (such as during transitions), leading to errors in the decision stream. Outlier detection schemes not only classify the data, but also verify that the current frame falls into the expected class distribution of that class using a one-class classifier (OCC). Various OCC algorithms have been used to detect outliers, including Support Vector Data Description (SVDD) \cite{liu_towards_2009, ding_adaptive_2019}, Gaussian Mixture Models (GMMs) \cite{ding_incremental_2022}, Hidden Markov Models \cite{kumar_verification-based_2021}, and Mahalanobis Distance (MD) \cite{ding_real-time_2017}. If $g_k(.)$ is an arbitrary OCC for class $k$ that returns $-1$ if the input feature is an outlier, otherwise returns a $+1$. The scheme is given by:    
        \begin{equation}
            \hat{y_i} = \begin{cases*}
                \hat{y}_i & if $\max\limits_k(g_k(X_i)) > 0$,\\
                \bigotimes & o.w.
            \end{cases*}
        \end{equation}        
        where $\bigotimes$ is the class that the decision will be assigned if the feature vector corresponding to the frame is an outlier. OD may reject out-of-domain patterns, but requires multiple one-class classifiers and may not be resilient to in-domain, but previously unseen patterns (such as due to confounding factors). \\
        
    \end{itemize}

\section{Newly Proposed Dynamic Rejection Schemes}

A major challenge with CBR-based schemes is how to determine the optimal rejection threshold ($\mathrm{Th_{Rej}}$). Setting a high value for $\mathrm{Th_{Rej}}$ will increase the rate at which incorrect decisions are rejected, but will also lead to an increase in rejecting correct decisions and vice-versa. In this work, we attempt to get the best of both worlds by introducing two new rejection schemes that vary the rejection threshold dynamically to react to changes in the decision stream. Our experience with continuous EMG suggests that most erroneous decisions have a locally volatile nature (particularly during transitions) motivating the need for a time-varying rejection threshold based on the temporal nature of the decision stream. In other words, if the decision stream is highly unstable, it might be prudent to set a high rejection threshold. The threshold can then be relaxed when the instability drops to avoid over-rejection in steady-state. Consequently, we propose the follow two new schemes:\\

\noindent \emph{Decision-Change Informed Rejection}: Decision-Change Informed Rejection (DCIR) is designed to react to changes in the decision stream. It models the determination of rejection threshold in a manner similar to that of the voltage across a capacitor: every time a change in decision is detected in the decision stream, (i.e. $\hat{y}_i \neq \hat{y}_{i-1}$), the threshold is set to a high value to reject all but very confident decisions. Over time, as the decisions stabilise to a single class, the threshold decays exponentially to avoid over-rejection. In effect, this scheme behaves like a high-pass filter where high frequency changes in decisions are penalized heavily. The rejection threshold for frame $i$, $\mathrm{Th_{Rej}}_i$ is determined by:
\begin{equation}
    \mathrm{Th_{Rej}}_i = \mathrm{Th}_{\min} + (\mathrm{Th}_{\max} - \mathrm{Th}_{\min})\times \exp^{\frac{-l}{\tau}}
\end{equation}

where $\mathrm{Th}_{\min}$ represents the minimum rejection threshold, $\mathrm{Th}_{\max}$ represents the maximum rejection threshold, $\tau$ represents the time constant, and $l \geq 0$ represents the number of frames since the last decision change.\\ 

\noindent \emph{Variance of Confidence Informed Rejection}: Variance of Confidence-Based Rejection (VoCIR) is designed to react to changes in the confidence stream. This scheme simply sets the rejection threshold  proportionally to the variability in the confidence stream: a large variability in the confidence stream corresponds to large uncertainties and/or fluctuations in decisions, and therefore the threshold is set to a high value to reject all but very confident rejections. The rejection threshold for frame $i$, $Th_{{Rej}_i}$ is determined by:
\begin{equation}
    \mathrm{Th_{Rej}}_i = \min(\mathrm{Th}_{\max}, \mathrm{Th}_{\min} + \beta \times \mathrm{v}_i)
\end{equation}

where $\mathrm{Th}_{\min}$ represents the minimum rejection threshold, $\mathrm{Th}_{\max}$ represents the maximum rejection threshold, $\beta$ represents the sensitivity, and $\mathrm{v}_i$ represents the maximum variance of the confidence stream across all the classes in the current and last $m$ frames, given by:
\begin{equation*}
    \mathrm{v}_i = \max_k(\sigma^2(c_{ik}, c_{(i - 1)k}, \ldots , c_{(i - m)k}))
\end{equation*}

where $\sigma^2(.)$ returns the variance.\\

\section{Methods}

\subsection{Data Acquisition}

Data was collected from forty three able-bodied participants (age: $25.98 \pm 5.8$, 26M, 17F), recruited mainly from a graduate student population. All participants gave informed consent. The study was approved by the University of New Brunswick’s Research Ethics Board (REB \#2021-116).

Surface EMG signals were collected from six bipolar electrodes placed around the circumference of the forearm, a third of the way down the forearm, proximal to the elbow. The electrodes were affixed in an equidistant clockwise fashion starting above the middle of the flexor carpi radialis muscles.  A Delsys Trigno System \cite{noauthor_trigno_nodate} was used to record sEMG signals which were sampled at \SI{2}{\kilo\hertz} with a 16-bit Analog-To-Digital converter. Additionally, a Leap Motion Controller \cite{noauthor_tracking_nodate} was used to simultaneously record the position of the hand to serve as a ground truth for identifying transition regions. The Leap Motion Controller samples position data at a variable rate, up to a maximum of \SI{120}{\hertz}. Both the Leap and EMG signals were recorded with a custom data collection program built using Python (version 3.10). The EMG signals were bandpass filtered between \qtyrange{20}{450}{\hertz} using a \nth{4} order zero-phase filter to remove any low or high frequency noise \cite{simao_review_2019, samuel_intelligent_2019}. 

\subsection{Training and Testing Protocol}
The data collection protocol used in this study was similar to the one used in our previous work, with some added elements \cite{tallam_puranam_raghu_analyzing_2022}. Each training record included a set of 6 ramp contractions starting from a neutral position and ending with a \SI{3}{\second} steady state contraction in Wrist Flexion (WF), Wrist Extension (WE), Wrist Pronation (WP), Wrist Supination (WS), Chuck Grip (CG), or Hand Open (HO). No Movement (NM) was also included yielding 7 classes per record.  Participants completed 5 trials guided by a visual computer prompt. Participants held each contraction for 3 seconds, and then returned to a neutral position for 3 seconds before being prompted to start the next contraction.After segmenting out the neutral periods, 5 repetitions (\SI{3}{\second} long) of each of the 7 classes per participant were used as training data.

Each test record included a continuous transition from each class to all others, generating a total of 7 x 6 = 42 transitions. Participants were guided by a visual computer prompt to move randomly from one contraction to another holding each in steady-state for \SI{3}{\second}. The motion leap position data was used to identify the bounds of transition regions based on a velocity calculated from the hand orientation vectors.  A threshold based on variability in the no movement data was used to accommodate random fluctuations.  Based on this identification process, 6 steady-state segments for each class were identified along with the 42 transitions in each record. Participants completed six trials.

\subsection{Comparing DSQI Algorithms}

\subsubsection{Classifiers}
\begin{figure}[htbp]
    \centering
    \includegraphics[width=0.45\textwidth]{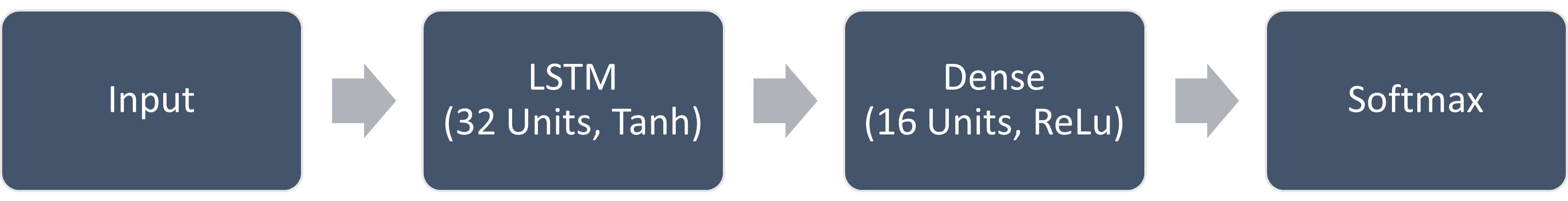}
    \caption{LSTM Network architecture. The number of units and the activation functions are shown in parentheses, as applicable.}
    \label{fig:LSTM_Model}
\end{figure}

The training and test records were segmented into overlapping frames with a length of $\SI{160}{\milli\second}$ and an increment of $\SI{16}{\milli\second}$. The Low-Sampling Frequency 4 (LSF4) feature set \cite{phinyomark_feature_2018, campbell_current_2020} was extracted from each frame, as this feature set has been shown to robust and generalizable. Three classifiers were trained with the ramp training data features: a Linear Discriminant Analysis (LDA) classifier, a Support Vector Machine (SVM) classifier, and a Long Short-Term Memory (LSTM) classifier. The classifiers were chosen to represent generative, discriminative, and deep learning models, respectively, and have been shown to perform well in offline EMG PR analysis\cite{zhang_adaptation_2013, robertson_effects_2019, jabbari_emg-based_2020}. LDA was implemented as described in \cite{scheme_confidence-based_2013}, SVM was implemented using Scikit-Learn (ver 1.1.2) \cite{pedregosa_scikit-learn:_2011}, and LSTM was implemented using Keras (ver 2.10) \cite{noauthor_keras:_nodate}. The hyperparameters of SVM and LSTM were tuned using a grid-search method and data from five randomly chosen participants using $10$-fold cross-validation using only the ramp training data. Since five participants represent only a small subset of dataset, this ensures that hyperparameters were not overtuned to this specific dataset. For the SVM, we varied the $C$ and $\gamma$ values in steps, and for the LSTM, we varied the number of units. Based on those pilot results, a single value was chosen for each hyper-parameter and used for the remainder of the experiment. For the SVM, the chosen hyperparameters were: $C=100$, RBF Kernel with $\gamma=0.01$; and for the LSTM network, the architecture is shown in Figure \ref{fig:LSTM_Model}. The Adam optimizer was used with a learning rate of $\SI{2e-4}{}$, the batch size was set to $64$, and the network was trained for $100$ epochs with early stopping to prevent over-fitting.

\begin{figure*}[b!]
    \centering
    \includegraphics[width=\textwidth]{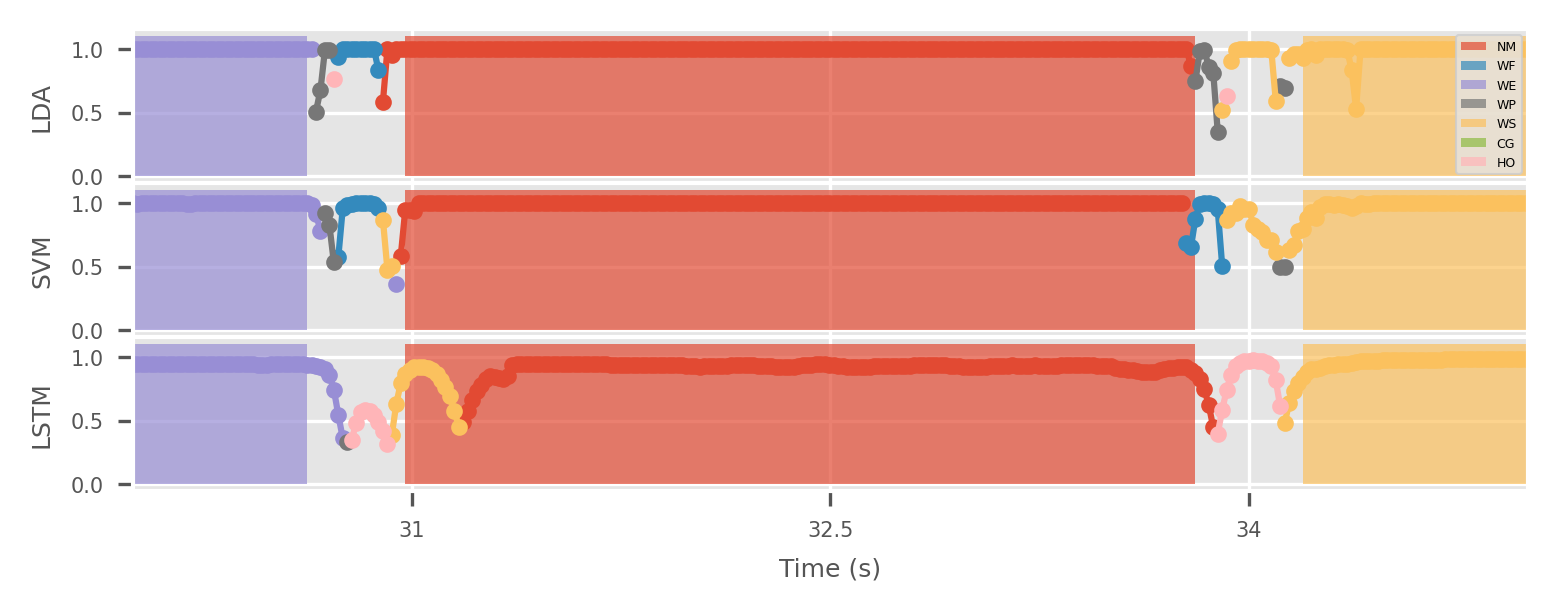}
    \caption{Decision stream comparison of the three classifiers (LDA, SVM, and LSTM) without any post-processing scheme. The 5 seconds shown include two transition zones representative of typical behaviour. Colours of the data points represent the classifiers decision (as indicated in the legend), and values represent confidence; shaded regions denote the ground truth class (based on the Leap data).}
    \label{fig:TS_0}
\end{figure*}

\begin{figure*}[t!]
    \centering
    \includegraphics[width=\textwidth]{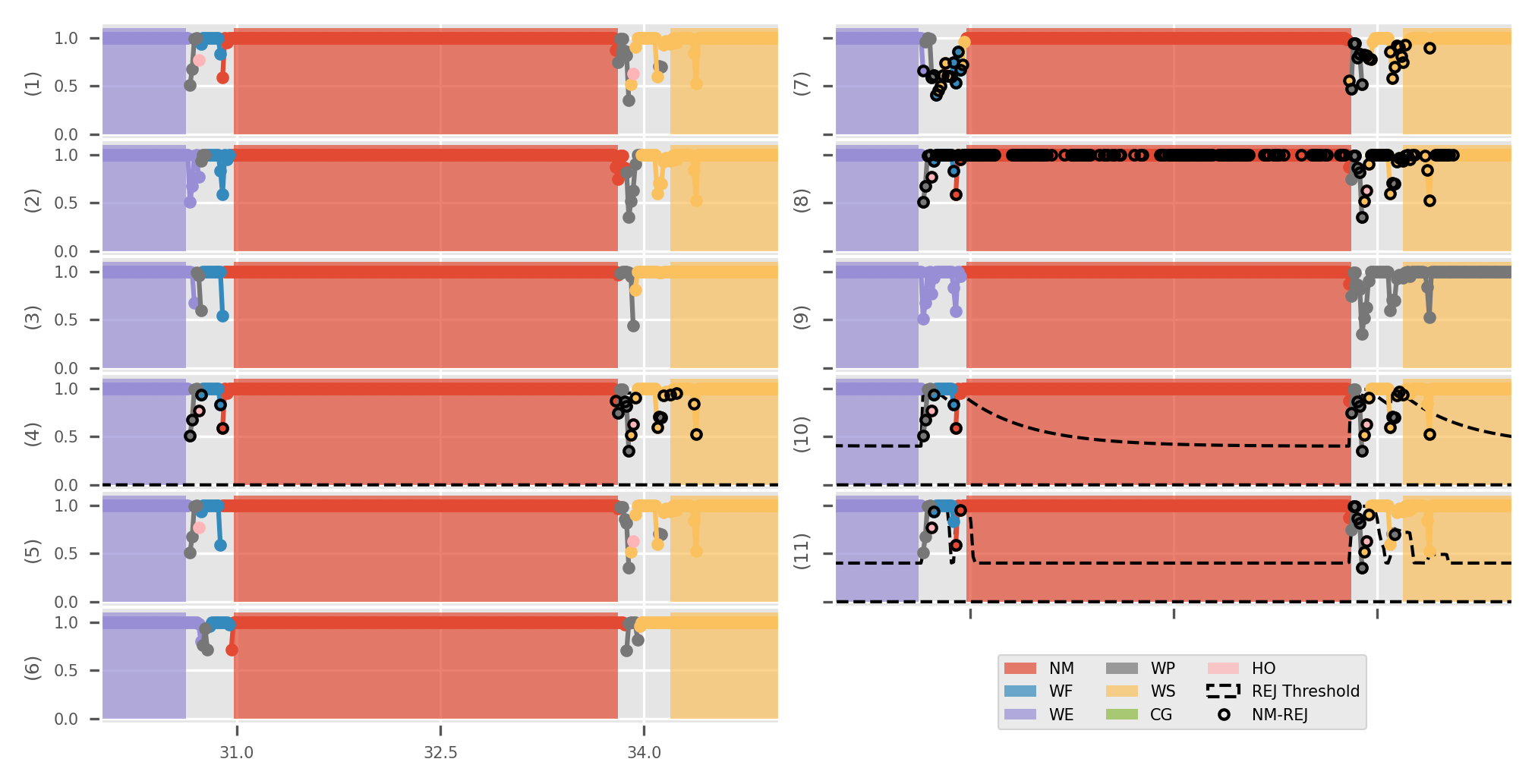}
    \caption{Comparison of the different post-processing schemes with an LDA decision stream. The 5 seconds shown include two transition zones representative of typical behaviour for LDA. Colours of the data points represent the classifiers decision (as indicated in the legend), and values represent confidence; shaded regions denote the ground truth class (based on the Leap data); and black circles denote frames that are rejected by the post-processing scheme. (1): No Post-processing, (2): MV, (3): PLDA, (4): CBR, (5): CS, (6): BF, (7): AW, (8): OD, (9): OL, (10): DCIR, (11): VoCIR.}
    \label{fig:TS_1}
\end{figure*}

\begin{table}[!t]
    \begin{center}
    \caption{Summary of DSQI Schemes. Dec = Decision-Based, Conf = Confidence-Based, Feat = Feature-Based \label{tbl:summary_dsqi_schemes}}
    \begin{tabular}{p{0.1\columnwidth}|p{0.08\columnwidth}|p{0.7\columnwidth}}
        \hline
        \textbf{Scheme} & \textbf{Family} & \textbf{Description}\\
        \hline
        MV & Dec & Use the most frequent class of the past $m$ decisions as the new estimate\\
        pLDA & Dec & Increase the weight of the last observed class by increasing its prior probability\\
        CBR & Conf & Accept decisions with high confidence values, reject to no movement otherwise\\
        CS & Conf & Increase the weight of the no movement class by scaling the confidences\\
        BF & Conf & Smooth confidence by taking the weighted average of the last $m$ confidences, then find the class with the highest smoothed confidence\\
        AW & Conf & If confidence is low, reject to no movement, increase the frame length, and try again\\
        OL & Feat & Reject frames below a low amplitude threshold to no movement, lock class after onset until contraction falls below threshold again \\
        OD & Feat & Reject to no movement frames with features outside expected distributions\\
        \hline
        \multicolumn{3}{c}{\textbf{Proposed Techniques}}\\
        \hline
        DCIR & Dec & Increase rejection threshold during rapid decision fluctuations, then gradually decay the threshold to the specified minimum\\
        VoCIR & Conf & Increase rejection threshold proportional to the observed variability in confidence of the past $m$ decisions\\

        \hline
    \end{tabular}
    \end{center}
    
\end{table}


\subsubsection{DSQI Hyperparameters}
The following hyperparameters were set for the DSQI algorithms:
\begin{itemize}
    \item \textbf{MV}: We used $m=8$ past values based on empirical testing.
    \item \textbf{pLDA}: We set $P_{\max} = 0.97$ and $b = 0.5$ based on the results of a grid search method.
    \item \textbf{CBR}: The rejection threshold for LDA was set to $\mathrm{Th_{Rej}} = 0.97$ based on \cite{campbell_linear_2019}. For SVM, it was set to $\mathrm{Th_{Rej}} = 0.6$ based on \cite{robertson_effects_2019}, and for LSTM, it was set to $\mathrm{Th_{Rej}} = 0.6$ based on a grid search.
    \item \textbf{CS}: We set the scaling factor for NM class to be $0.97$, and all the others to be $0.05$, based on the original study.
    \item \textbf{BF}: We used $m=8$ past values, as with MV.
    \item \textbf{AW}: We varied the frame length from the default of $\SI{160}{\milli\second}$ to $\SI{256}{\milli\second}$ in increments of $\SI{16}{\milli\second}$. The rejection threshold for LDA was set to $\mathrm{Th_{AW}} = 0.97$ based on \cite{campbell_linear_2019}, and to $\mathrm{Th_{AW}} = 0.6$ for SVM based on \cite{robertson_effects_2019} AW was not used with the LSTM as it is was unclear how it could be implemented with a temporal classifier.
    \item \textbf{OL}: We set $m=6$, as in the original study.
    \item \textbf{OD}: We used the SVDD to reject outlier frames to the NM class ($k_\mathrm{NM}$).
    \item \textbf{DCIR}: We set $\mathrm{Th}_{\min} = 0.4$, $\mathrm{Th}_{\max} = 0.989$, and $\tau = 30$. These were tuned to provide comparable steady-state rejection to CBR.
    \item \textbf{VoCIR}: We used $m=8$ past values, similar to MV. We set $\mathrm{Th}_{\min} = 0.4$, $\mathrm{Th}_{\max} = 0.989$, and $\beta = 4.0$. Similar to DCIR, these were tuned to provide comparable steady-state rejection to CBR.
    
\end{itemize}

\subsubsection{Assessing Performance}
We evaluated the 30 scheme-classifier combinations using 3 steady-state region metrics  - Active Error Rate (AER), Total Error Rate (TER), and Instability (INS) along with 6 transition region metrics - Offset Delay ($T_{OFFSET}$), Onset Delay ($T_{ONSET}$), Transition Duration ($T_{TRANSITION}$), Instability (INS), Tertiary Class Error (TCE), and Percent No Movement (PNM), as described in our previous work \cite{tallam_puranam_raghu_analyzing_2022}. To calculate the transition metrics, it was necessary to identify when each classifier moved in and out of steady-states. This was done for each steady-state segment by comparing the known state class (based on the prompt) with the classifier decision stream in the region of a transition bound (as identified with the leap data). The first decision beyond a transition entry bound that did not correspond to the known state was considered to be associated with steady-state offset. Likewise, the first decision beyond a transition exit bound that did correspond to the known state was considered to be associated with steady-state onset. Majority vote was used across 9 frames in both cases to handle blips in the decision stream.

For each participant, the various metrics were first averaged across steady-states and across transitions within a continuous test trial, then across the six test trials. The results were analysed for significant differences using a One-Way Analysis of Variance (ANOVA). When significant differences were indicated, we performed a paired t-test (with Holm correction) to identify the groups that differ from each other. For all cases, $\alpha=0.05$ was used to denote significance. The statistical analysis was done using the \textit{statsmodels} (ver 0.13.2) \cite{seabold_statsmodels:_2010} and \textit{scikit-posthocs} (ver 0.7.0) \cite{terpilowski_scikit-posthocs:_2019} Python packages, and the visualisation was done using the \textit{matplotlib} (ver 3.6.1) \cite{hunter_matplotlib:_2007}, \textit{Pandas} 
\cite{mckinney-proc-scipy-2010}
and \textit{Seaborn} (ver 0.12.0) \cite{waskom_seaborn:_2021} Python packages.

\section{Results}

\begin{figure*}[b!]
    \centering
    \includegraphics[width=\textwidth]{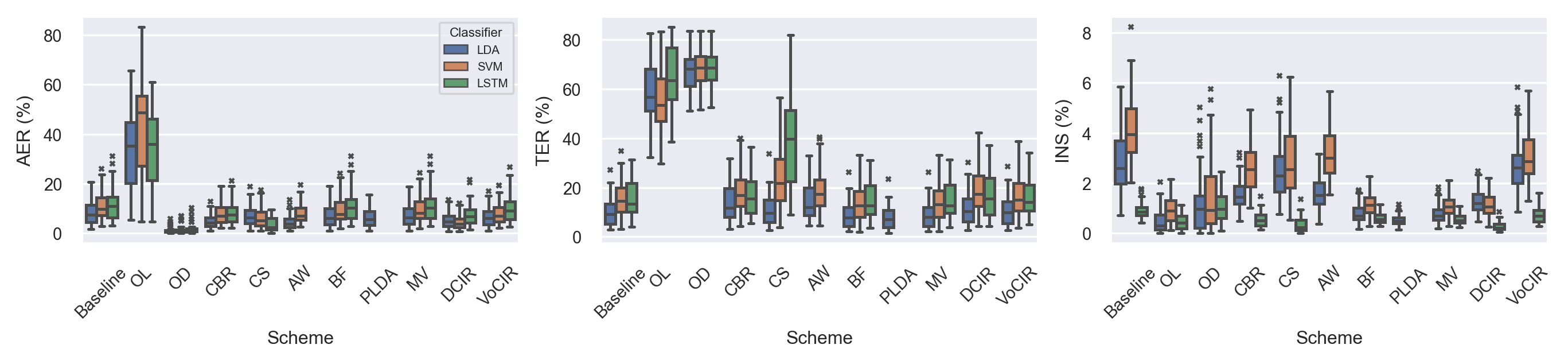}
    \caption{Box-plots of steady-state metrics across all classifiers and schemes.}
    \label{fig:BP_SS}
\end{figure*}

Our goal was to compare different DSQI schemes in conjunction with different types of classifiers in the context of transitions as well as steady-states and to highlight the trade-offs made by the schemes. A visual depiction of the confidence streams obtained from the three classifiers for an example set of transitions is shown in Figure \ref{fig:TS_0}. The plots show colour-coded confidence streams, where the color of each decision point denotes the classifier-assigned label, and the ground truth labels derived from the Leap sensor as color-shaded regions. The plots exemplify the typical confidence distribution and decision stream characteristics observed across the three classifiers. A further comparison of the effects of each of the post-processing scheme is shown in the subplots \qtyrange{1}{11}{} in Figure \ref{fig:TS_1}.

\begin{figure*}[t!]
    \centering
    \includegraphics[width=\textwidth]{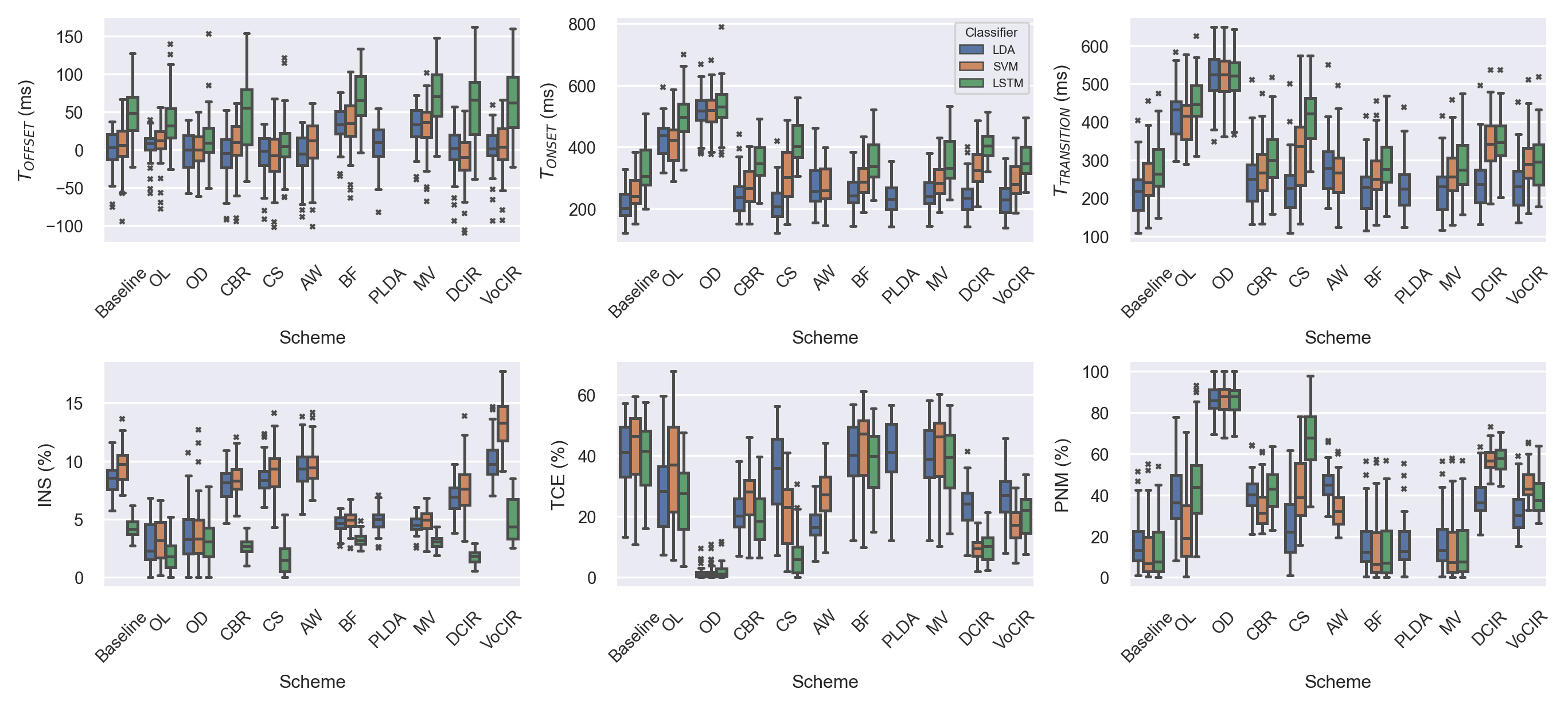}
    \caption{Box-plots of transition metrics across all classifiers and schemes.}
    \label{fig:BP_TR}
\end{figure*}

Figures \ref{fig:BP_SS} and \ref{fig:BP_TR} depict box plots of the observed steady-state and transition metrics, respectively, for the various the DSQI schemes tested. 

Results of the ANOVA test suggest a statistically significant difference across the schemes for all steady-state and transition metrics. Post-hoc tests reveal that without any post-processing, LDA, SVM, and LSTM do not exhibit statistically different steady-state AER and TER. However, all three classifiers were statistically different from each other when considering INS, with the LSTM exhibiting substantially smaller INS than LDA or SVM ($p < 0.0001$). The LDA had the second best steady-state INS performance followed by the SVM, which had the worst INS performance across all cases. 

The rejection schemes tended to trade a reduction in AER for an increase in TER, as seen in previous works \cite{scheme_comparison_2015, robertson_effects_2019}, and explained in \cite{Scheme_Englehart_Hudgins_2011}. Though Post-hoc analysis indicates that OD had the best AER, it also revealed that the OD and OL schemes exhibit substantially poorer performance in TER; both schemes saw a $>40\%$ increase in TER, a seemingly unfavourable trade-off. As a result, for the rest of the analysis, OD and OL are excluded from the analysis.
For the remaining schemes, LDA tended to yield lower average AER than SVM and LSTM, although no statistically significant differences were noted. 

All of the rejection schemes reduced steady-state INS. However, when applied with LDA and SVM, the MV and BF schemes could do no better than the \emph{baseline} LSTM ($p > 0.9$ for LDA, and $p > 0.3$ for SVM). Moreover, the addition of MV or BF to the LSTM reduced the steady-state INS further, making it significantly better than the baseline LSTM ($p < 0.0005$), and thus, outperforming both LDA and SVM with those schemes.  Nevertheless, the newly proposed DCIR with LSTM was found to be the best combination in terms of steady-state INS, and was revealed to be statistically different than all other schemes ($p < 0.001$) except for CS with LSTM.

The transition metrics indicate that without any post-processing, LDA and SVM had similar transition delay characteristics for $T_{OFFSET}$, $T_{ONSET}$, and $T_{TRANSITION}$, indicating that these two classifiers take similar amount of time to transition between any two classes ($p > 0.9$, $p > 0.1$, and $p > 0.9$ respectively). However, the LSTM had statistically slower responsiveness and transition time than LDA and SVM ($p < 0.005$), indicated by its larger $T_{OFFSET}$, $T_{ONSET}$, and $T_{TRANSITION}$ values. The LSTM also tended to have the lowest transition INS among the classifiers, exhibiting similar trends to the steady-state metrics. Generally, most DSQI schemes had similar impact on the delay metrics, however, the largest increase in delay was seen in CS with the LSTM, suggesting that this combination may yield potential over-rejection.

Only rejection-based schemes appear to have impacted TCE and PNM, which may be expected as they are able to suppress uncertain decisions. The best classifier-scheme combination was the new LSTM with DCIR, which produced significantly lower TCE than all of schemes except CS with LSTM ($p < 0.0001$). 

\section{Discussions}

In this work, we explored the performance trade-offs of 8 previously established post-processing algorithms and 2 novel approaches on EMG-based pattern recognition decision stream quality. We used our recently proposed framework to show how these algorithms perform during steady-state as well as in the context of transitions, which are substantially more challenging and contribute disproportionately to overall classifier error. In general, we observed that the newly proposed decision-change informed rejection (DCIR) approach, which dynamically adjusts rejection thresholds based on decision stream volatility, is a promising algorithm that reduced transition errors and INS metrics. It performed substantially better than confidence-based rejection (CBR) during transitions with comparable steady-state TER, indicating that the addition of temporal information into conventional rejection systems is beneficial. With this added context, however, the new schemes trade off these performance gains with a slight decrease in responsiveness of the classifiers. Nevertheless, this trade-off is still an improvement over approaches like majority vote (MV), which also decreases responsiveness with no improvements in TCE.

Based on the increase in steady-state TER, the onset locking (OL) and outlier detection (OD) schemes were the worst performers in combination with any classifier. Both of these schemes make assumptions that only work with well-behaved data, which may explain the results under our more challenging test conditions. OL assumes that all movements are followed by a period of rest (i.e., NM class) \cite{zhang_improving_2017}. This may be reasonable in very specific circumstances, such as gesture recognition, but is generally not true in myoelectric applications. The researchers in \cite{zhang_improving_2017} acknowledged that their scheme may break down under dynamic conditions, and we were indeed able to demonstrate this in our test conditions. In our test dataset, the participants transitioned from one class to another without necessarily resting in between, which led to the OL scheme locking onto a class but not unlocking after a transition. This phenomenon can be observed in the decision stream plot for OL in Figure \ref{fig:TS_1} (9). This results in the substantial increase in TER, which can be seen in the box plots in Figure \ref{fig:BP_SS}.

Similarly, the OD schemes assume that the distribution of test data are perfectly consistent with those observed in training conditions. Though the scheme may reject frames from contractions that are not part of the training set \cite{liu_towards_2009}, it may also incorrectly reject frames from contractions that \emph{are} part of the training set due to factors such as co-variate shift.  Even the ramp data used to train the classifiers is more well-behaved than the data seen in the continuous test dataset, resulting in the scheme over-rejecting a substantial amount of test data as `outliers'.

The confidence scaling (CS) scheme was promising in conjunction with the LSTM, though in this case, it appears that the scaling was overly aggressive. The weights across the different classifiers were set based on reported work \cite{campbell_linear_2019}, but our results suggest that these could be tweaked further in the case of the LSTM and SVM to possibly achieve a better trade-off between different metrics. Further exploration for optimization is warranted, but caution should be taken to avoid over-tuning parameters for any one dataset.

The adaptive windowing (AW) scheme appears to provide very limited gains in performance while also being one of the more computationally demanding algorithms. AW requires that the classifiers be trained with features extracted from all possible frame lengths \cite{al-timemy_adaptive_2018}, which leads to a substantial increase in processing time. In addition, it is unclear how the temporal schemes such as the LSTM should be trained for this scheme, as the temporal dynamics change with frame length, which does not affect the LDA or SVM, but significantly affects the LSTM. The original work compared AW with MV and Bayesian fusion (BF) schemes, but not rejection-based schemes. We found AW to be arguably better than MV and BF, in agreement with the original study, but CBR, DCIR, and VoCIR all provide comparable or better performance across the metrics while being significantly less computationally demanding.

The LSTM classifier had the least baseline instability, though it also took slightly longer to transition to a new class. LSTM is a temporal classifier and takes into account the past frames along with the current frame when making a decision. Therefore, it has additional context, and is able to suppress blips in the decision stream. However, the training data used to train classifiers in myoelectric control is well-behaved and does not contain dynamics associated with transitions, and therefore, the LSTM struggles to understand when it should be transitioning into (or out of) a class. Our empirical testing suggests that training an LSTM with data containing transitions may allow the LSTM to learn these dynamics and transition faster; however the user burden associated with collecting all such transitions for training may be prohibitive.

The rejection schemes were overall better than MV and BF smoothing schemes. They not only reduced TCE (by increasing PNM), but also slightly improved INS. In contrast, MV and BF only served to reduce INS (in line with previous results \cite{englehart_robust_2003}), without tangibly affecting other metrics. DCIR was found to be the most promising in this regard, as it significantly increases PNM (and consequently reduces TCE) without also correspondingly increasing steady-state TER.

Although LDA was the most promising baseline classifier, in line with findings from our previous study \cite{tallam_puranam_raghu_analyzing_2022}, it does not benefit as much from the rejection schemes, particularly DCIR and VoCIR, as compared to the LSTM and SVM. This can be attributed to the significant overlap in confidence value distributions of the LDA corresponding to the different types of decisions (correct, incorrect, and transitions). In other words, the LDA is often too confident, even with the wrong decisions. This leads to the undesirable situation where it is not possible to reject the wrong decisions without also rejecting many correct decisions. However, in the case of LSTM and SVM, the confidence values of the different types of decisions have a smaller overlap, allowing the rejection schemes to make much better use of the temporal characteristics of the confidence streams. With DCIR, both the LSTM and SVM saw a $>40\%$ increase in PNM, resulting in a much lower TCE, and consequently, a reduction in unwanted motions during transitions. 

It is currently unclear how these techniques may interact with other approaches to myoelectric control, such as regression \cite{Smith_Kuiken_Hargrove_2016}. Regression is promising as it may enable the simultaneous independent control of multiple degrees of freedom; however, it requires a more controlled training protocol than classification and suffers from many similar challenges as classification based approaches (e.g. inadvertent activation of tertiary classes during transitions). Consequently, more work is required to fully understand how DSQI approaches such as DCIR may be adapted to benefit regression control.

\section{Conclusion}
Existing work \cite{robertson_effects_2019} indicates that errors during transitions greatly degrade usability, which supports the inclusion of transition metrics as part of offline classification comparison studies. Evaluating classifier schemes with continuous test data enables transition metrics to be considered along with steady state ones and using this approach, we demonstrated that the LSTM-DCIR combination is a strong offline performer. Nevertheless, it is not yet clear how to weight the importance of each of the metrics from a usability perspective and our results clearly indicate that trade-offs are necessary; for instance, the LSTM-DCIR trades reduced responsiveness for improved tertiary errors counts. Studies have reported that prosthesis users have noted inadvertent or unwanted activation as being destructive to their sense of agency \cite{chadwell_characterisation_2022}, and that users prefer over-rejection over responsiveness \cite{scheme_comparison_2015}. These findings suggest that TCE may be an important metric, but further research is warranted to explore the relationship between offline steady-state and transition metrics and online performance directly, through a usability study. 

\printbibliography
\end{document}